\documentclass[aps,prc,twocolumn,superscriptaddress,showpacs,preprintnumbers,amsmath,amssymb]{revtex4}
\usepackage{graphicx}% Include figure files

\begin{document}

%\vspace*{1.5cm}
\def	\capoor	{\rm^{40}Ca}
\def	\ca	{\rm^{48}Ca}
\def	\kr	{\rm^{80}Kr}
\def	\zbound	{Z_{\rm bound}}
\def	\zmax	{Z_{\rm max}}

%% Definitions of units %%
\def	\ns	{{\rm ns}}
\def	\cm	{ {\rm cm}}
\def	\MeV	{ {\rm MeV}}
\def	\AMeV	{ {\rm AMeV}}
\def	\MeVc	{ {\rm MeV/c}}
\def	\AMeVc	{ {\rm AMeV/c}}

%% definitions of numeric constants %%
\def	\elab	{E_{\rm lab}}	%LAB energy

\def\KONYA{Department of Physics, University of Sel\c{c}uk, TR-42079 Konya, Turkey}
\def\GSI{GSI Helmholtzzentrum f\"{u}r Schwerionenforschung GmbH, D-64291 Darmstadt, Germany}
\def\ITP{Institute for Theoretical Physics, J. W. Goethe University, D-60438 Frankfurt am Main, Germany}
\def\HFHF{Helmholtz Research Academy Hesse for FAIR (HFHF), Max-von-Laue-Str. 12, D-60438 Frankfurt am Main, Germany}

\newcommand{\goo}{\,\raisebox{-.5ex}{$\stackrel{>}{\scriptstyle\sim}$}\,}
\newcommand{\loo}{\,\raisebox{-.5ex}{$\stackrel{<}{\scriptstyle\sim}$}\,}

%\title{Isospin influence on isotope yields in nucleus collisions around the Fermi Energy}
\title{Isospin compositions of correlated sources in the Fermi energy domain}

\affiliation{\KONYA}
\affiliation{\GSI}
\affiliation{\ITP} 
\affiliation{\HFHF}
%\affiliation{\NIC}

%\affiliation{\FIAS}

\author{R.~Ogul}        	\affiliation{\KONYA}\affiliation{\GSI}  
\author{A.~S.~Botvina}       	\affiliation{\ITP}\affiliation{\HFHF}
\author{M.~Bleicher}            \affiliation{\GSI}\affiliation{\ITP}\affiliation{\HFHF}%\affiliation{\NIC}  
\author{N.~Buyukcizmeci}    	\affiliation{\KONYA}
\author{A.~Ergun}	    	\affiliation{\KONYA}
\author{H.~Imal}	    	\affiliation{\KONYA}
\author{Y.~Leifels}             \affiliation{\GSI}
\author{W.~Trautmann}           \affiliation{\GSI}

\date{\today}% It is always \today, today,
             %  but any date may be explicitly specified

\begin{abstract}
Isotopic yield distributions of nuclei produced in peripheral collisions of $^{80}$Kr+$^{40,48}$Ca at 35 MeV/nucleon 
are studied. Experimental results obtained by the FAZIA Collaboration at the LNS facility in Catania are compared
with calculations performed with the statistical multifragmentation model (SMM).  
The fragments with atomic number $Z=19-24$ observed at forward angles are successfully 
described with the ensemble method previously established for reactions at higher energy. Using the SMM results, the isotopic compositions 
of the projectile residues are reconstructed.
The results indicate a significant isospin exchange between the projectile and target nuclei,
not far from isospin equilibrium, during the initial phase of the reaction. 
%The subsequent statistical decay of the excited sources is found to be responsible for the minor target dependence of the final fragment
%distributions observed experimentally.  
The two groups of light fragments with $Z=1-4$, experimentally distinguished by their velocities relative to coincident heavy projectile 
fragments, are found to originate from 
different sources. The isotopic composition of the slower group is consistent with emission from a low-density neck, enriched in neutrons,
and satisfactorily reproduced with SMM calculations for a corresponding neck source of small mass.
\end{abstract}


%\keywords{Suggested keywords}%Use showkeys class option if keyword
                              %display desired
\maketitle

\section{INTRODUCTION}
\label{sec:int}

There is continuing high interest in investigating the  
nuclear fragment production in heavy-ion collisions of intermediate 
energies~\cite{dyntherm,nusyme}. It is a typical many-body process in which the nuclear 
interaction can manifest its ingredients and lead to important 
states of matter. The fragment production mechanisms, the properties 
of nuclear matter at both high and low densities, new
reaction products, e.g., exotic isotopes, and their correlations can be 
addressed~\cite{nusyme,exotic}. The theoretical explanations of these phenomena 
include various methods involving dynamical and statistical approaches. 
In such complicated many-body processes, phenomena governed by interactions of both, high and low energy 
coexist. Their description may go beyond the capabilities 
of one specific model. It is, therefore, important to 
determine correctly the regions of applicability for each model and to explore
alternatives to reach a comprehensive picture.

%FAZIA
The technical achievements of the FAZIA Collaboration have widened the possibilities for reaction studies 
at intermediate energies~\cite{FAZIA21,camaiani21,ciampi22}. 
By extending the limits of mass and charge resolution of telescope-type detection units, 
arrays with large solid-angle coverage have become available for new experiments exploring the role
of isospin in heavy-ion collisions~\cite{bougault14,pastore17}. 
Isospin transfer between the reaction partners reflects the action of the symmetry parts of the nuclear forces~\cite{nusyme,baran05,lipr08}.
Studying their effects requires isotopic identification of the reaction products.
It has been known for a long time that product distributions respond to the isotopic properties of the primary 
source, whose composition, however, may be concealed by secondary-decay effects~\cite{porile64,kukarol77}.

In a first experiment addressing isospin transport at intermediate energies, 
the FAZIA Collaboration studied the fragment production in $\kr$ collisions with $\ca$ and
$\capoor$ targets at 35 MeV/nucleon beam energy~\cite{FAZIA21}. 
Four detector cubes equipped with the FAZIA technology were placed at forward angles.
The chosen inverse kinematics had the effect of focusing projectile fragments into 
forward directions efficiently covered by the FAZIA modules. 

The analysis of the data was aimed at extracting the density dependence of the 
symmetry-energy term in the nuclear equation of state. For this 
purpose, the antisymmetrized molecular dynamics (AMD) transport model and the statistical decay code
GEMINI were used and ``a weak indication in favor of a stiff symmetry energy" was reported. 
At densities below saturation, the stiff solution corresponds to lower values for the symmetry energy.
However, comparison with the experimental data shows that the isotopic composition 
of the observed fragments is not much influenced by the symmetry-energy term and 
depends mainly on the statistical processes at the later reaction stages. 
For this reason, we believe it is instructive to perform an alternative analysis of 
the experimental results in order to clarify the role of the statistical breakup and decay in the formation of the final products. 
%In GEMINI++, the Hauser-Feshbach formalism is implemented for the n, p, d, t, 3He, a, 6He, 6–8Li, and 7–10Be channels.

The present work attempts an interpretation of the same data aiming at the isospin transfer during the 
contact time between the reaction partners. The isotopic composition of the primary reaction products 
will be determined from the observed products by modeling the deexcitation of the former with the statistical multifragmentation model (SMM, Ref.~\cite{SMM}). 
The SMM is capable of following various reaction scenarios from compound evaporation through fission and multifragmentation up to vaporization
according to the statistical weights given to these paths in the model description. It thus represents a rather universal model 
which was previously successfully applied for the description of the fragment production in peripheral and central heavy ion collisions at 
relativistic energies~\cite{Bot95,Xi97,Ogul11,Imal15} and in the Fermi energy domain (Refs.~\cite{MSU,INDRA,TAMU,Bot06} and references given therein). 

To describe the projectile fragments, the    
calculations start from an ensemble of excited sources whose properties, mainly the isotopic composition, is varied
until an optimum description of the measured isotopic distributions for projectile fragments with atomic numbers $Z=19-24$ is achieved.
The peripheral character of the data has encouraged us to attempt an extrapolation of the method found useful for reactions 
at higher energies~\cite{Bot95,Ogul11} to the present case. It differs from an earlier theoretical study of peripheral $^{112}$Sn + $^{124}$Sn 
and $^{124}$Sn + $^{112}$Sn reactions 
in which sources with fixed mass and charge were used as initial configurations for SMM calculations~\cite{Imal20}.

Normalized relative yields of light fragments from the same experiment are also presented in the FAZIA publication~\cite{FAZIA21}.
They were recorded in coincidence with a heavy fragments of $Z \ge 12$ ($Z > 18$ in the case of $\kr$ + $\capoor$)
and sorted into two groups depending on whether their longitudinal velocities 
are larger or smaller than those of the coincident heavy fragment. Considering that fragments with $Z = 3$ or 4 included in the faster group 
are not typical evaporation products, their origin is not immediately obvious. The slower group extends up to $Z = 11$ and thus covers the $Z$ interval 
of intermediate mass fragments abundantly produced in multifragmentation processes. Their neutron content $\langle N \rangle /Z$ as a function of
$Z$ exhibits the odd-even staggering known from other reactions, revealing a neutron richness exceeding that of the projectile sources.
To explain these fragments we assume the formation of a highly excited statistical source forming a neck between the target and the 
projectile~\cite{ditoro06}.
With SMM calculations based on this scenario, the observed isotopic composition
is very satisfactorily reproduced.

\section{FRAGMENT PRODUCTION IN THE STATISTICAL APPROACH}

Presently there are two main statistical conceptions intending to describe the 
physics of fragment production as a decay of excited finite nuclei. The most 
popular one is that of the compound nucleus \cite{Bohr} which decays with the sequential emission 
of light particles. There are several mathematical formalisms according to this conception 
which can be used for practical calculations. For example, GEMINI~\cite{charity88,charity90} uses the 
Hauser-Feshbach method, while SMM uses the Weisskopf method for the low-energy emission. 
As was shown in the theoretical and experimental studies, the compound nucleus conception works very 
well from low excitation energy of nuclei up to values around 2 to 3 MeV/nucleon~\cite{SMM,pienkowski02}. 

Another conception was mainly developed in the 1980s (see, e.g., reviews 
\cite{SMM,Gross90} and references therein), and we may call it the freeze-out-volume 
conception. It was motivated by intensive experimental studies of the multifragmentation 
process leading to the production of many intermediate mass fragments from the 
decay of a highly excited nuclear source. 

According to the freeze-out-volume 
conception, the compound nucleus has no time to be formed, since the highly excited 
nucleus rapidly expands (on a time scale of approximately 100 fm/$c$), and the system breaks-up 
(reforms itself) into fragments during this expansion. The last moment at which the fragments can 
still be considered as interacting and being under formation is the statistical freeze-out 
time. Subsequently the fragments propagate in the mutual Coulomb fields and undergo 
secondary deexcitation as usual for low-excited nuclei. SMM includes this conception 
and thus provides an opportunity to describe multifragmentation data. 
Since we are dealing with finite nuclear systems, a weak point of the 
freeze-out-volume conception up to now was that the upper limit of the excitation 
energy was not determined. In this case, the very idea of applying statistical laws in the rapidly 
expanding matter is not convincing from a theoretical view. 

A method for solving this problem was suggested recently: By analyzing the fragment production 
in central nucleus-nucleus collisions at beam energies of more than 50--100 MeV/nucleon, it was established that 
the maximum excitation energy of such statistical sources can not reach more than 
8--10 MeV/nucleon~\cite{Bot21,Bot22}. This compares well with the previous maximum excitation 
energy of sources found in peripheral and central collisions. It is consistent 
with the binding energy of finite nuclei, providing the natural physical limit for 
the application of the statistical theory. In our case of nucleus-nucleus collisions at the Fermi 
energy, such energies correspond to the maximum energy reached in the center of mass 
of the colliding nuclei. 

According to the statistical hypothesis, initial dynamical 
interactions between the colliding nuclei lead to a redistribution of the 
available energy among many degrees of freedom. The nuclear system 
evolves towards equilibrium. In a general consideration, the 
process may be subdivided into two stages: (1) a dynamical stage 
leading to the formation of an equilibrated nuclear system, and 
(2) the statistical disassembly of the system into individual primary 
fragments, followed by their deexcitation and the formation of the final product distribution observed in experiments. 

Dynamical models indicate that individual reactions at intermediate energies 
are not evolving as equilibrium processes~\cite{furuta09}. However, because of the complexity 
of the dynamical processes, the intermediate reaction states populate a phase space
that can be described with very few parameters, and whose disassembly according to statistical laws
leads to the observed asymptotic state of the reaction~\cite{raduta07}. By applying the SMM, a scenario is adopted 
that places the equilibrium state at an early stage of the reaction, before the partitioning
of the excited system occurs. The parameters describing the equilibrium state thus contain information on
global properties of the initial reaction stages forming it. The present study aims at identifying 
the isospin transport during this early part of the collision.

Sequences of statistical decay processes are expected to proceed towards the
valley of stability at which binding energies reach their maximum. More precisely, the expected asymptotic destination 
is represented by the evaporation attractor line (EAL), whose location is mainly determined
by the competition between proton and neutron evaporation~\cite{charity98}. Often, and also in the present case, 
the decay sequences are not long enough, so that the EAL is approached but not fully reached.

The emission of fragments is treated in the SMM by expanding the excited sources 
to the statistical freeze-out volume. This is assumed to be caused by thermal pressure or 
as a decompression process after an initial dynamical collective compression. 
However, a similar low-density freeze-out state of nuclear matter may also be
created through the stochastic knock-out of nucleons produced by
initial nucleon-nucleon collisions. The freeze-out density is expected to be around (0.1--0.3)$\rho_0$, where 
$\rho_0 \approx 0.16$~fm$^{-3}$ is the normal nuclear density. The value of 0.3$\rho_0$ has been adopted for the
projectile sources as in previous analyses \cite{Bot95,Xi97,Ogul11}.

Peripheral reactions at the present intermediate energies are additionally characterized by a low-density third source of nucleons and light
fragments located at intermediate velocities~\cite{ditoro06,montoya94,dempsey96,lukasik97,piantelli02,baran04,hudan04}.
It may contain products of nucleon-nucleon collisions removed from the main residues as well as remnants of necklike structures 
joining projectile and target residues before they separate. 
Interpreted as a continuation of the multifragmentation mechanism towards 
intermediate impact parameters, the processes at intermediate velocity seem suitable for a statistical description with the 
SMM, an approach rarely attempted up to now~\cite{Erg15}. 
The neutron content of these sources is of specific interest
because of the expected neutron drift toward regions of lower density caused by the density dependence of the symmetry energy~\cite{Bot01,Colonna05}.

\section{Method}

In Ref.~\cite{FAZIA21}, the FAZIA Collaboration reports the relative isotope distributions for elements with
$Z=19-24$, individually normalized for each atomic number $Z$. 
They were measured with four FAZIA detector blocks placed at forward angles
between $\theta_{\rm lab} = 2.4^{\circ}$ and $17.4^{\circ}$, covering approximately 14\% of the solid angle 
up to $\theta_{\rm lab} = 17.4^{\circ}$ (see Ref.~\cite{FAZIA20} for a layout of the experiment). 
In the present analysis, we assume that the fragment distributions recorded within the 
covered solid angle are representative for projectile fragmentation in  the studied reactions. The same is assumed for the
coincident light fragments with atomic numbers $Z \le 4$ and $Z \le 11$ recorded with the same setup for two velocity regimes.

The velocities of the $Z=19-24$ fragments are shown to be only slightly 
lower than the initial velocity of the $\kr$ projectiles. This, together with their masses of more than 
half of the projectile mass, identifies them as resulting from peripheral processes.
One may thus expect that their masses are correlated and their excitation energies inversely correlated 
with the impact parameter. Such correlations are reminiscent of the participant-spectator scenario established 
for reactions at higher energies. It is, therefore, further assumed that the equilibrated intermediate state can be 
described with an ensemble of sources with properties similar to those determined in the analysis of projectile fragmentation at energies 
up to 1 GeV/nucleon~\cite{Bot95,Xi97,Ogul11,Imal15}. 

\begin{figure} [tbh]
\centerline{\includegraphics[width=8.0cm]{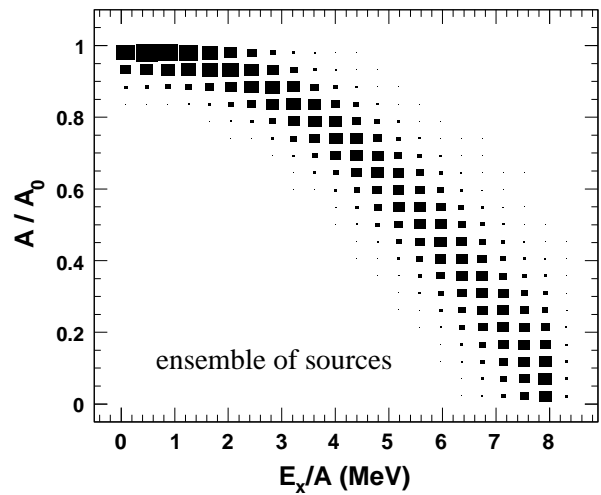}} 
%\centerline{\includegraphics[width=8.0cm]{figures_riza/Figure5b_ensemble.eps}} 
\caption{\small{Ensemble of hot thermal sources represented in a scatter plot of reduced mass
number $A/A_0$ versus excitation energy $E_x/A$, as used in the SMM calculations. 
The frequency of the individual sources is proportional to the area of the squares 
(reprinted with permission from Ref.~\protect\cite{Ogul11}; copyright \copyright~2011 by
the American Physical Society).
}}
\label{fig:ensemble}
\end{figure}

The ensemble consists of a distribution of sources extending from systems of low excitation and masses near the projectile mass to 
highly excited systems of small mass expected to be formed in central collisions, and, in this form, is universally applicable (Fig.~\ref{fig:ensemble}).
The form of its peripheral branch with excitation energies up to $\approx 4$~MeV/nucleon compares well with AMD predictions for comparable 
reactions~\cite{hudan06} and the general energy-mass correlation is consistent with 
transport model predictions up to much higher energies~\cite{Bot17}.  %WT the cited paper reports on 5 GeV/nucleon energies
The fragments of interest and
their distributions are selected from the output configurations obtained from the eventwise performed calculations for the full ensemble.

\vspace{4 mm}
 % fig. 1

Altogether, seven parameters are required to define the ensemble of excited sources. In the analyses performed with the ALADIN data for projectile
fragmentation at energies up to 1 GeV/nucleon, the experimental data permitted a determination of these parameters with very little 
ambiguity as shown in Refs.~\cite{Bot95,Ogul11}. This is not possible here, and the general form of the ensemble was, therefore, adopted from the earlier work.
The maximum mass $A_0=80$ and atomic number $Z_0=36$ are those of the projectile, assuming that pickup processes are rare at the present energies. 
Also the maximum excitation energy $E_x/A = 8$~MeV was chosen as in previous studies and as supported by recent work~\cite{Bot21,Bot22}. 
It is not crucial, however, because the event selection requiring a fragment with $Z \ge 12$ and $Z \ge 19$ for the two reactions implies that
the excited projectile residues compose at least about half of the original projectile.
The excitation energies of such sources with $A/A_0 \ge 0.5$ are limited to values near or below 4 MeV/nucleon (Fig.~\ref{fig:ensemble}).
As in previous studies (see, e.g., Refs.~\cite{MSU,INDRA,TAMU,Bot06,Imal20}), angular momentum is not considered as an independent 
property of the source because its effects are small and, regarding the isotopic fragment distributions, indistinguishable from the much larger
effects of the excitation energy.

\vspace{4 mm}
\begin{figure} [tbh]
\centerline{\includegraphics[width=8.0cm]{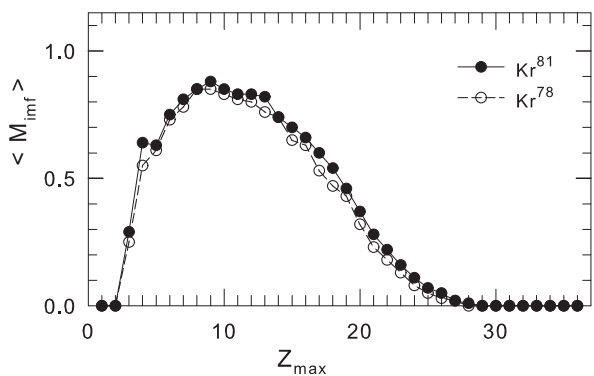}} 
%\centerline{\includegraphics[width=8.0cm]{figures_riza/Mimf-Zmax-modified.eps}} 
\caption{\small{Mean multiplicity $\langle M_{\rm imf} \rangle$ of fragments of intermediate mass with $Z \ge 3$ in coincidence with the fragment with the largest 
atomic number $Z_{\rm max}$ found in an event as a function of $Z_{\rm max}$, 
as obtained with the SMM for the fragmentation of the projectile in $^{81}$Kr + $^{48}$Ca (solid circles) and $^{78}$Kr + $^{48}$Ca (open circles). 
}}
\label{fig:mimf_zmax}
\end{figure} % fig. 2

The experimental data~\cite{FAZIA21} were collected without a particular
impact-parameter selection, even though peripheral collisions are favored by the detection at forward angles. 
Correspondingly, the whole ensemble of excited projectile 
residues is taken into account in the analysis, and the element range of interest is selected from the calculated set
of final products. 
For exploring the isospin transfer during the initial reaction stage, 
the mean composition $\langle N/Z \rangle$ was varied by varying the maximum mass $A_0$ without changing $Z_0$. It is assumed
that the associated small change in $E_x/A$ vs $A$, caused by the construction of the ensemble, is negligible for the narrow interval
of fragments with $Z = 19-24$. 

The isotope yields depend also on the properties of the 
primary hot nuclei in the statistical freeze-out. In this respect the most 
important ingredient is the symmetry term of their liquid-drop description and its coefficient $\gamma$. 
The symmetry term governs the widths of isobaric or isotopic fragment distributions~\cite{botvina85,Botvina02}. 
The standard value used in SMM applications is $\gamma \approx 25$~MeV~\cite{SMM,Buy05}. 
Similar but also lower values have been determined from statistical interpretations of reaction data at lower~\cite{souliotis07,hudan09} and
higher energy. The isoscaling analysis of fragmentation reactions with Sn targets and light-ion beams of up to 15 GeV incident energy resulted in
$\gamma = 22.5$~MeV~\cite{Botvina02}. From the isoscaling analyses performed for $^{12}$C + $^{197}$Au reactions at 300 and 600 MeV/nucleon~\cite{LeFevre05} 
and for the fragmentation of $^{107,124}$Sn and $^{124}$La projectiles on Sn targets at 600 MeV/nucleon~\cite{Ogul11} values of $\gamma = 15$~MeV and lower
were deduced. The lowest values, however, were found for the most violent collisions and for fragments with $Z \le 10$. 
A similar scenario is not expected in the present case for the production of fragments with $Z = 19-24$ in peripheral collisions at 35~MeV/nucleon.
As will be shown below, $\gamma$ can be determined rather precisely from the widths of the measured isotope distributions.

In a first step, the calculations were used to confirm that the chosen procedure permits satisfactory reproductions of
the experimental isotope distributions and to obtain global properties of the reaction. As an example, the multiplicity of fragments of
intermediate mass produced in coincidence with a largest fragment with atomic number $\zmax$ is shown in Fig.~\ref{fig:mimf_zmax}. The chosen
threshold is $Z \ge 3$ and the upper limit is either $\zmax$ or $36 - \zmax$, whichever is smaller. Fragments in this interval are not typical
for evaporation but, according to the calculations, are produced with mean multiplicities close to 1 for $\zmax$ around 10. 
In the interval covered by the experiment, $\zmax \ge 12$ and $\zmax \ge 19$ for the two reactions, the mean multiplicity of
associated fragments decreases from about 0.8 to very small values at the upper end of the range of $\zmax$. The multiplicity widths are probably large
and the production of intermediate mass fragments cannot be expected to be negligible.

In a second step, the sensitivity of the mean mass number of the produced projectile fragments to the isotopic compositions
of the chosen ensembles was explored. 
The compositions reveal the results of isospin transport phenomena during the initial phases of the collision, 
the main goal of the present study. The capabilities of the SMM~\cite{SMM} to describe the statistical disassembly 
of the intermediate system
and the subsequent modes of deexcitation permits a view back on to these early reaction stages through backtracing from the detected
fragmentation patterns.
In a third step, the origin of the coincident light fragments was investigated by comparing their mean neutron-over-proton ratios to
the results obtained for $^{107}$Sn and $^{124}$Sn fragmentations in experiment S254 conducted at the GSI laboratory~\cite{Ogul11}.

\vspace{1 mm}
\begin{figure} [tbh]
\centerline{\includegraphics[width=8.5cm]{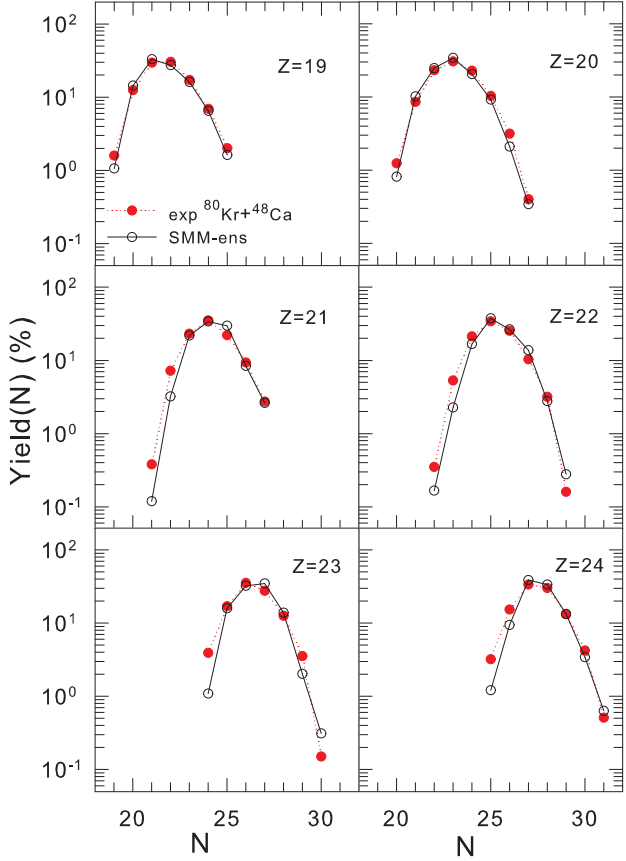}} 
%\centerline{\includegraphics[width=8.5cm]{figures_riza/Piantelli_Figure1.eps}} 
\caption{\small{Comparison of the predicted isotopic distributions of 
fragments with $Z=19-24$ with the experimental data measured in the 
$^{80}$Kr + $^{48}$Ca reaction. The solid red circles represent the experimental data 
from Ref.~\cite{FAZIA21} and open circles the results of the SMM-ensemble  
calculations for the same reaction. 
}}
\label{fig:1924_48}
\end{figure} % fig. 3

\vspace{1 mm}
\begin{figure} [tbh]
\centerline{\includegraphics[width=8.5cm]{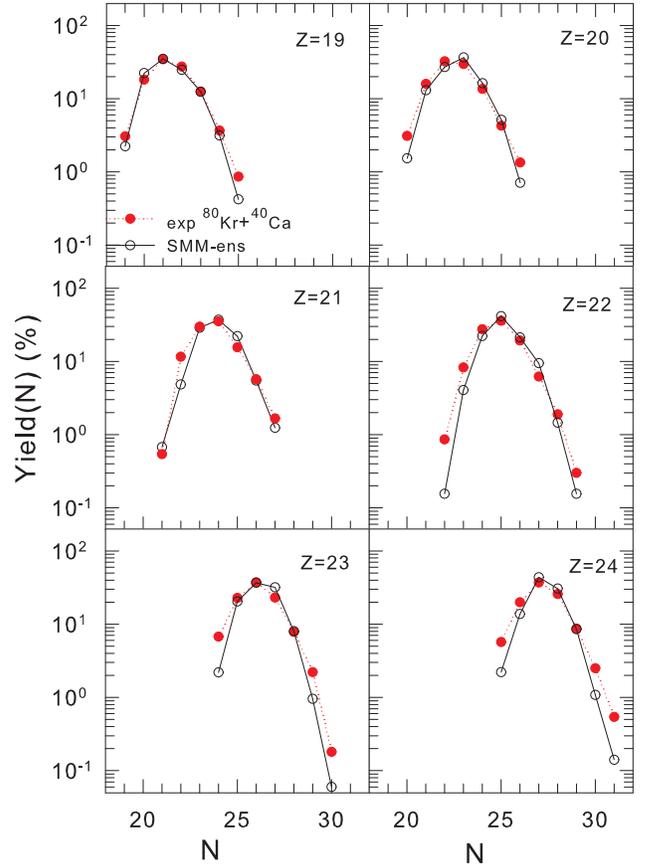}} 
%\centerline{\includegraphics[width=8.5cm]{figures_riza/Piantelli_Figure2.eps}} 
\caption{\small{Comparison of the predicted isotopic distributions of 
fragments with $Z=19-24$ with the experimental data measured in the 
$^{80}$Kr + $^{40}$Ca reaction. The solid red circles represent the experimental data 
from Ref.~\cite{FAZIA21} and open circles the results of the SMM-ensemble  
calculations for the same reaction. 
}}
\label{fig:1924_40}
\end{figure} % fig. 4

\section{Comparison with the experiment}

\begin{figure} [tbh]
\centerline{\includegraphics[width=8cm]{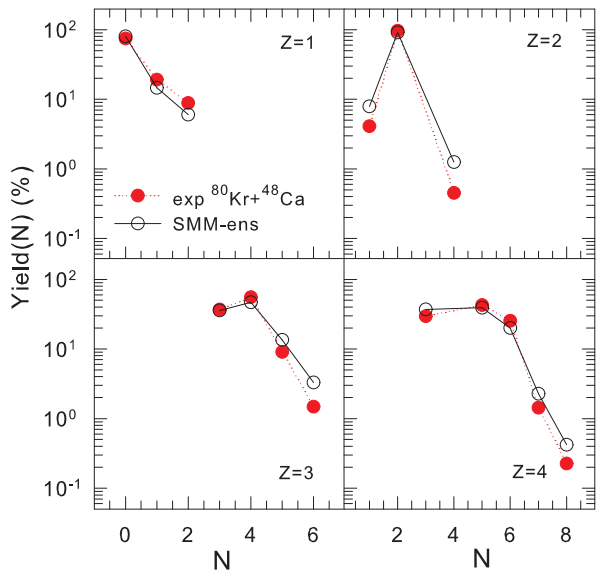}}     
%\centerline{\includegraphics[width=8cm]{figures_riza/Fig.5-IEVP2-gate12-36.eps}}     
\caption{\small{Isotopic yield distributions for light nuclei with atomic number $Z=1-4$, emitted forward with respect to the coincident
projectile residue in the $^{80}$Kr + $^{48}$Ca reaction. 
Red solid circles represent the experimental data of Ref.~\protect\cite{FAZIA21}, 
and open circles represent the results of the SMM-ensemble calculations.
}}
\label{fig:14_48}
\end{figure}

\begin{figure} [tbh]
\centerline{\includegraphics[width=8cm]{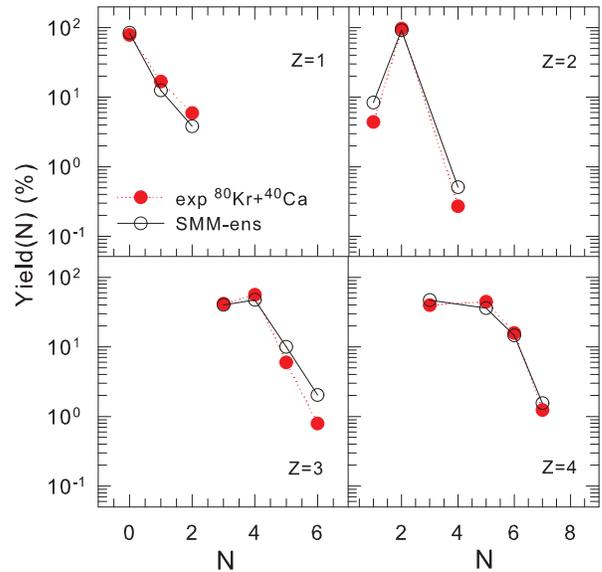}}    
%\centerline{\includegraphics[width=8cm]{figures_riza/Fig.6-IEVP2-gate19-36.eps}}    
\caption{\small{Isotopic yield distributions for light nuclei with atomic number $Z=1-4$, emitted forward with respect to the coincident
projectile residue in the $^{80}$Kr + $^{40}$Ca reaction. 
Red solid circles represent the experimental data of Ref.~\protect\cite{FAZIA21}, 
and open circles represent the results of the SMM-ensemble calculations.
}}
\label{fig:14_40}
\end{figure}

The results obtained in the first step are presented in Figs.~\ref{fig:1924_48} to~\ref{fig:14_40}. 
The isotopic distributions of nuclei in the interval of atomic numbers
$Z=19-24$ for collisions of 
$^{80}$Kr + $^{48}$Ca and $^{80}$Kr + $^{40}$Ca are shown in Figs.~\ref{fig:1924_48} and~\ref{fig:1924_40}, respectively.
The normalized isotopic yield distributions measured in the FAZIA experiments~\cite{FAZIA21} are represented by solid
circles. For the SMM calculations represented by the open circles, it was assumed that the projectile 
$^{80}$Kr picks up an extra neutron from the $^{48}$Ca target and loses two neutrons in interactions with the
$^{40}$Ca target, or that an equivalent exchange of nucleons takes place. With the corresponding parameters for the maximum
mass of the projectile $A_0 = 81$ and 78 and with $Z_0 = 36$, average compositions $\langle N/Z \rangle = 1.25$ and 1.17 were
thus assumed for the two cases, respectively. 

%\vspace{2 mm}
 % fig. 5

Other parameters describing the ensemble of sources were fixed as in Ref.~\cite{Ogul11}. 
Specifically, for the coefficient $\gamma$ of the symmetry term, values of $\gamma=18$~MeV for $Z=19-20$,
$\gamma=19$~MeV for $Z=21-22$, and $\gamma=20$~MeV for $Z=23-24$ were found
to provide the results closest to the experimental data for $^{80}$Kr + $^{48}$Ca and values larger by 1 MeV for $^{80}$Kr + $^{40}$Ca. 
The figures demonstrate that the chosen procedure 
permits a rather satisfactory reproduction of the experimental isotope distributions for the products of mid-peripheral collisions with
atomic number $Z=19-24$.
%from Program subroutine RESNUC: a1=0.001; a2=0.015; sig=0.07; c0=2

%\vspace{2 mm}
 % fig. 6

\begin{figure} [tbh]
\centerline{\includegraphics[width=8.5cm]{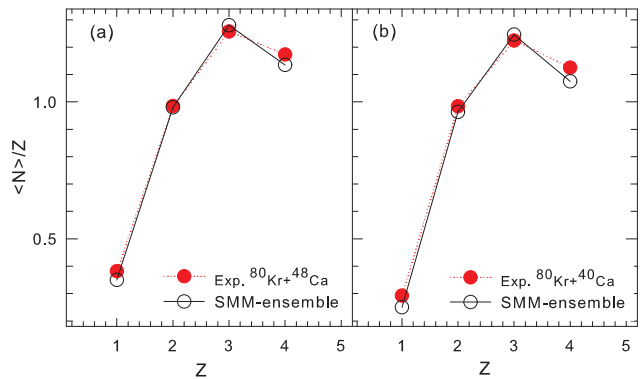}}   
%\centerline{\includegraphics[width=8.5cm]{figures_riza/Fig.8-IEVP2-Gate12-36and19-36.eps}}   
\caption{\small{Experimental and theoretical results for the
$\langle N \rangle/Z$ ratios of light fragments, emitted forward with respect to the coincident projectile residue,
obtained in the $^{80}$Kr + $^{48}$Ca 
and $^{80}$Kr + $^{40}$Ca reactions as presented in Figs.~\ref{fig:14_48} and \ref{fig:14_40}. 
Red solid circles represent the experimental data and open circles represent the results of the SMM-ensemble 
calculations. 
}}
\label{fig:14_noverz}
\end{figure} % fig. 7

%paragraph introducing Z=1-4 spectra
The same ensemble nuclei were also used to calculate isotopic distributions of light fragments with $Z=1 - 4$. The distinction made in the
experiment between particles emitted into forward or backward directions with respect to the coincident heavier fragment cannot be applied
in the model calculations but the coincidence with a heavier fragment was required. In the case of $^{80}$Kr + $^{48}$Ca, only events 
containing a fragment with atomic number $Z$ between 12 and 36 and, for $^{80}$Kr + $^{40}$Ca, only events containing 
a fragment with $Z$ between 19 and 36 were selected. 

Figures~\ref{fig:14_48} and \ref{fig:14_40} show the calculated relative isotopic yields of light nuclei with $Z=1-4$ in comparison
with the experimental yields for fragments with longitudinal velocities larger than those of the coincident heavier fragment.
%emitted after the decay of the primary sources in both reactions. 
Protons and $\alpha$ particles obviously dominate the distributions of hydrogen and helium isotopes.
In both reactions, $^7$Li is the strongest lithium isotope while $^9$Be is the strongest beryllium isotope only 
in the $^{80}$Kr + $^{48}$Ca reaction and is equally abundant as $^7$Be in the $^{80}$Kr + $^{40}$Ca reaction.
Overall, the observed differences between the two reactions appear to be small.
The results of the SMM-ensemble calculations represent the experimental data rather well.
On a logarithmic scale, deviations are essentially only visible in the cases of helium and lithium isotopes. Some of the predictions
for weak isotopes slightly overestimate the experimental yields. 
These yields play a minor role for the mean neutron-over-proton ratios shown for the $Z=1-4$ products in Fig.~\ref{fig:14_noverz}.
The agreement is best for $Z=2$, obviously so because of the dominance of $^4$He, and overall very satisfactory.

\section{Symmetry coefficient $\gamma$}

Within the SMM, the statistical disintegration of highly excited nuclei includes the expansion of the nuclear system into a 
freeze-out volume within which the primary hot fragments are formed and the chemical equilibrium (i.e., the detailed-balance 
interaction) between them is established. 
The properties of these hot fragments can be different from those of cold stable nuclei. In particular, their symmetry-term coefficients 
$\gamma$ can be lower than the standard value of 25 MeV that is taken in the liquid-drop Bethe-Weizs\"{a}cker formula describing masses of normal 
nuclei (see, e.g., Refs.~\cite{SMM,Botvina02,Buy05,souliotis07,hudan09,LeFevre05}). 
The reason is that the fragments are interacting until freeze-out, they can be slightly expanded, and their surface parts can be 
extended. Subsequently, the standard value of the symmetry-term coefficient 
is restored during the secondary deexcitation of the hot fragments~\cite{Buy05}. 

\begin{figure} [tbh]
\centerline{\includegraphics[width=8.5cm]{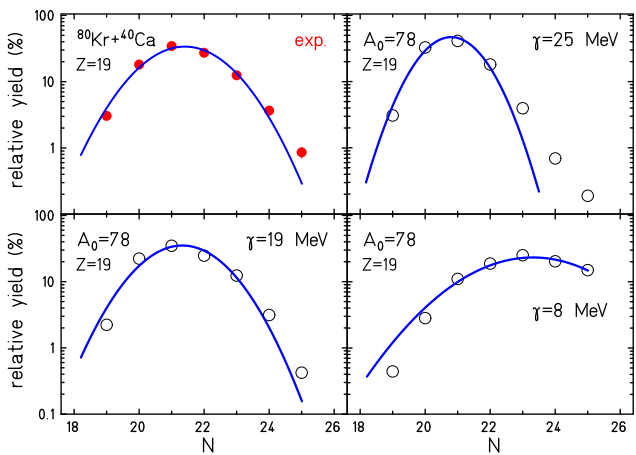}}
%\centerline{\includegraphics[width=8.5cm]{gauss_all_19_4.eps}}
\caption{\small{Relative yield distributions for $Z=19$ and the results of Gaussian fits (blue lines) 
as measured in $^{80}$Kr + $^{40}$Ca (top left panel, red solid circles) and as calculated with $A_0=78$ and the indicated values of the symmetry-term
coefficient $\gamma$ (open circles). The errors of the experimental data are mostly smaller than the symbol size.
}}
\label{fig:gauss4}
\end{figure}

\begin{figure} [tbh]
\centerline{\includegraphics[width=6.5cm]{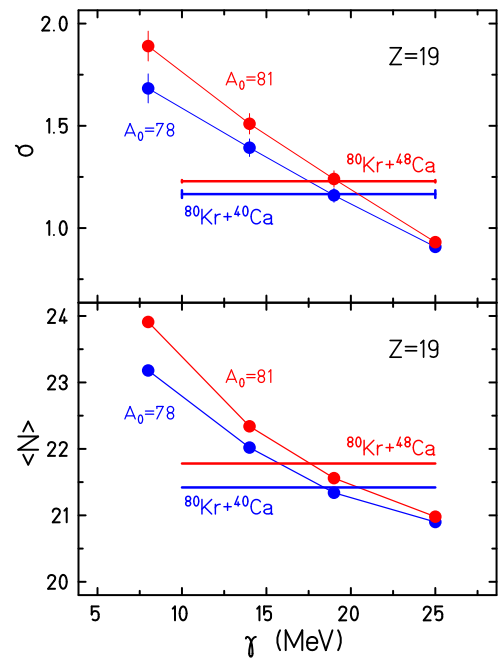}}
%\centerline{\includegraphics[width=6.5cm]{gauss_check7.eps}}
\caption{\small{The standard deviation $\sigma$ of the Gaussian functions fitted to the measured and calculated isotope distributions 
in units of the neutron number (top) and the mean neutron number
$\langle N \rangle$ (bottom) as a function of the symmetry-term coefficient $\gamma$ used in calculations for $Z = 19$ with $A_0 = 78$ and 81 (filled circles). 
The horizontal lines indicate the results deduced from the experimental data including their uncertainties which, for the case of $\langle N \rangle$,
are smaller than the line widths. 
}}
\label{fig:sigma}
\end{figure}

The coefficient $\gamma$ 
controls the widths of the calculated isotope distributions~\cite{botvina85,Botvina02}. 
This is illustrated in Fig.~\ref{fig:gauss4}, which shows the examples of 
isotopic yields for $Z=19$ calculated for $A_0 = 78$ and values of $\gamma = 8$, 19, and 25~MeV in comparison with the experimental data
obtained in $^{80}$Kr + $^{40}$Ca. It is evident that the widths change significantly with
$\gamma$ and that the value $\gamma=19$~MeV chosen for this case leads to the good reproduction of the data seen in Fig.~\ref{fig:1924_40}. 
It is also evident that the Gaussian fits included in the figure permit a quantitative analysis of the distribution widths. The dependence of the Gaussian
standard deviations on $\gamma$ is illustrated in the top panel of Fig.~\ref{fig:sigma}. It confirms that $\gamma=19$~MeV for $Z=19$ 
is rather precisely, with an uncertainty of around $\pm 1$~MeV, 
determined by the measured data. 
The dependence of the mean neutron number $\langle N \rangle$ on $\gamma$, equally
evident in Fig.~\ref{fig:gauss4}, is given in the bottom panel of Fig.~\ref{fig:sigma}. 

The widths of some of the isotope distributions appear to be 
slightly underpredicted by the calculations (Figs.~\ref{fig:1924_48} and~\ref{fig:1924_40}). An adjustment of $\gamma$ in these cases requires the 
simultaneous modification of the mean $\langle N \rangle /Z$ that is to be expected. The variations of the widths and mean neutron numbers 
with $\gamma$, shown for $Z=19$ in Fig.~\ref{fig:sigma}, are rather universal and very similar for the studied elements with $Z=19-24$. Further 
individual optimizations of specific parameters may lead to a slightly different ensemble of excited sources representing the intermediate
reaction systems.

 % fig. 8

 % fig. 9

The values $\gamma = 18-21$~MeV applied in the present calculations had already been found useful for describing projectile fragments 
of similar magnitude from $^{112,124}$Sn + $^{112,124}$Sn collisions at 1~GeV/nucleon~\cite{Imal15}. 
They are also close to the $Z$ dependent symmetry-term coefficients used in the microcanonical multifragmentation model~\cite{raduta07b} 
and in the AMD model~\cite{ono04}.
The surface term included in the AMD leads to $\gamma = 20.4$~MeV and 21.2~MeV for $Z = 19$ and 24, respectively, assuming $A = 2Z$. The AMD
coefficients were determined by fitting the ground state binding energies for $A \le 40$ nuclei calculated with the AMD model. 
In the present case, the symmetry term 
is part of the description of the produced nuclear fragments with excitation energies up to about 4 MeV/nucleon. 
The difference of the coefficients is, however, marginal.

\section{Initial compositions}

The initial compositions defined by choosing $A_0 = 78$ and 81 indicate a considerable amount of isotopic equilibration between projectile 
and target during the early stages of the two reactions, more than 60\% of full isospin equilibrium. For a more quantitative analysis,
the $N/Z$ compositions of the ensembles of excited sources were investigated by 
varying the $A_0$ parameter from 75 to 83 while keeping $Z_0 = 36$ fixed. It covers the interval of initial
compositions $\langle N/Z\rangle_{\rm ini} = A_0/36 -1$ from 1.083 to 1.306.
The subscript "ini" is used here and in the following to indicate the initial isotopic composition 
at the beginning of the statistical decay modeled with the SMM.
%The simultaneous variation of the coefficient $\gamma$ of the symmetry term within its range of uncertainty of about $\pm 1$~MeV
%has no visible effect on the isotope distributions that are obtained.

\begin{figure} [tbh]
\centerline{\includegraphics[width=8.0cm]{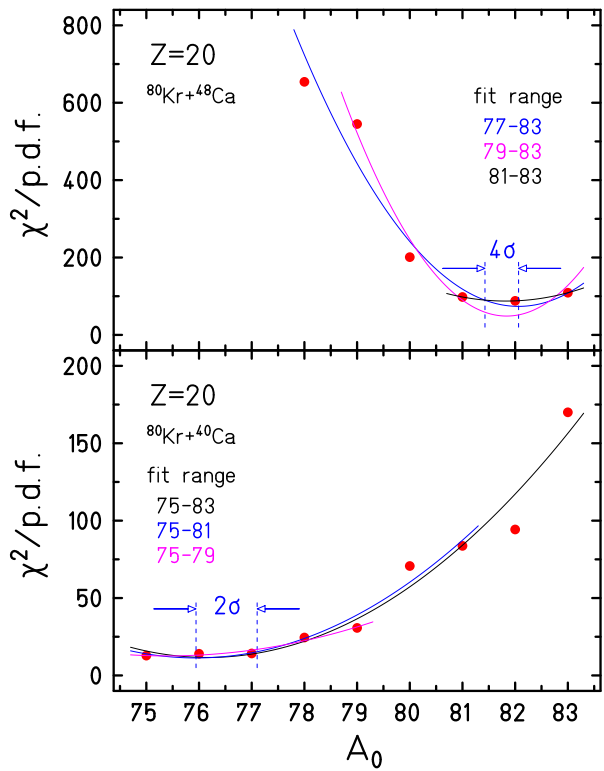}}
%\centerline{\includegraphics[width=8.0cm]{chi2_20a.eps}}
\caption{\small{Mean quadratic deviations ($\chi^2$ per degree of freedom) of the relative isotopic yields for $Z = 20$ 
projectile fragments with mass numbers $A = 40 - 46$ as a function of the $A_0$ parameter used for varying the initial composition
$\langle N/Z\rangle_{\rm ini}$ of the ensemble of sources for the two reactions $^{80}$Kr + $^{48}$Ca (top) and $^{80}$Kr + $^{40}$Ca (bottom). 
The lines are the results of parabolic fits performed for three different intervals of $A_0$ as indicated in each panel. The uncertainty intervals
obtained for the fits extending over seven data points (blue lines) corresponding to $4 \sigma$ in the upper and $2 \sigma$ in the lower panel,
are shown for illustration. 
}}
\label{fig:chi2}
\end{figure} % fig. 10

The calculated fragment yields were normalized, as in the experimental report~\cite{FAZIA21}, 
and the sum of the quadratic deviations from the measured values in units of the experimental errors was 
determined for each of the sets obtained with a given $A_0$. The so obtained $\chi^2$ distribution was divided by the number of 
isotopes considered for a given $Z$, reduced by one because of the normalization. The result for the example of $Z = 20$ fragments
is shown in Fig.~\ref{fig:chi2} for the two studied reactions. The displayed distributions of $\chi^2$ per degree of freedom exhibit the expected
parabolic behavior. Their absolute values are rather large because the quoted experimental errors are very small~\cite{FAZIA21},
considerably smaller for the data obtained with the $^{48}$Ca target than for the data obtained with the $^{40}$Ca target.

The minima of the parabolic fits (Fig.~\ref{fig:chi2}) determine the optimum composition of the ensemble of sources as approximately $N/Z_{\rm ini} = 1.28$, 
obtained with $A_0 = 82$, for the fragmentation on the $^{48}$Ca target and $N/Z_{\rm ini} = 1.11$ obtained with $A_0 = 76$ for the
$^{40}$Ca case. They are similar to the original assumptions but indicate an even larger degree of equilibration.
By analyzing the fits for the intervals extending over seven data points (represented by the blue lines), 
$A_0 = 77 - 83$ for $^{48}$Ca and $A_0 = 75 - 81$ for $^{40}$Ca, the compositions
$N/Z_{\rm ini} = 1.280 \pm 0.004$ and $N/Z_{\rm ini} = 1.110 \pm 0.016$ are obtained for the two cases, respectively. The corresponding 4$\sigma$ and
2$\sigma$ widths are indicated in the figure. 

\begin{table}
\caption{\label{tab:summary}
Summary of the results obtained for the initial configurations $\langle N/Z \rangle_{\rm ini}$ by comparing the calculated mean
neutron numbers $\langle N \rangle$ with the experimental value (fourth column) or by searching for the minimum of the
$\chi^2$ distribution obtained with the experimental errors (last column). The prediction of the evaporation attractor 
line~\protect\cite{charity98} is shown in the third column.
}
\begin{ruledtabular}
\begin{tabular}{l c c c c c c}
$^{48}$Ca & & & $\langle N/Z \rangle_{\rm ini}$ & $\langle N/Z \rangle_{\rm ini}$\\
\hline
$Z$ & $\langle N \rangle_{\rm exp}$ & $N/Z_{\rm EAL}$ & from $\langle N \rangle_{\rm exp}$ & from $\chi^2$ \\
\hline
\\
19 & $21.78\pm 0.01$ & 1.116 & $1.278\pm 0.003$ & $1.300\pm 0.005$ \\
20 & $23.11\pm 0.01$ & 1.118 & $1.292\pm 0.002$ & $1.280\pm 0.004$ \\
21 & $24.10\pm 0.01$ & 1.121 & $1.209\pm 0.002$ & $1.215\pm 0.007$ \\
22 & $25.23\pm 0.01$ & 1.123 & $1.206\pm 0.002$ & $1.198\pm 0.006$ \\
23 & $26.39\pm 0.01$ & 1.125 & $1.210\pm 0.002$ & $1.207\pm 0.007$ \\
24 & $27.49\pm 0.01$ & 1.128 & $1.220\pm 0.002$ & $1.244\pm 0.009$ \\
& weighted mean & & $1.232\pm 0.001$ & $1.254\pm 0.002$ \\
\\
\hline
$^{40}$Ca & & & $\langle N/Z \rangle_{\rm ini}$ & $\langle N/Z \rangle_{\rm ini}$\\
\hline
$Z$ & $\langle N \rangle_{\rm exp}$ & $N/Z_{\rm EAL}$ & from $\langle N \rangle_{\rm exp}$ & from $\chi^2$ \\
\hline
\\
19 & $21.42\pm 0.02$ & 1.116 & $1.207\pm 0.011$ & $1.169\pm 0.017$ \\
20 & $22.53\pm 0.02$ & 1.118 & $1.098\pm 0.008$ & $1.110\pm 0.016$ \\
21 & $23.77\pm 0.02$ & 1.121 & $1.067\pm 0.010$ & $1.126\pm 0.013$ \\
22 & $24.92\pm 0.02$ & 1.123 & $1.060\pm 0.011$ & $1.115\pm 0.014$ \\
23 & $26.10\pm 0.02$ & 1.125 & $1.072\pm 0.006$ & $1.113\pm 0.016$ \\
24 & $27.21\pm 0.02$ & 1.128 & $1.103\pm 0.007$ & $1.113\pm 0.023$ \\
& weighted mean & & $1.101\pm 0.003$ & $1.124\pm 0.006$ \\
\\
\end{tabular}
\end{ruledtabular}
\end{table}             %Table I

The results obtained for the six elements are given in the last column of Table~\ref{tab:summary}. 
The reconstructed intermediate configurations of the excited 
projectile systems are fairly neutron rich for $^{80}$Kr + $^{48}$Ca and fairly neutron poor for $^{80}$Kr + $^{40}$Ca. The weighted mean values are
1.254 and 1.124, respectively, with a difference much larger than naively expected from the fairly similar experimental isotope 
distributions (Figs.~\ref{fig:1924_48} and~\ref{fig:1924_40}; see also Fig.~6 in Ref.~\cite{FAZIA21}). 
It is obvious that the individual errors obtained for each element are not representative for the element-to-element variations. 
For further use, we therefore chose to give equal weight
to each of the elements for determining the mean values and to adopt the mean squared deviations from the mean divided by $\sqrt 5$ as their uncertainties. 
With this choice, the compositions obtained with the $\chi^2$ method are $\langle N/Z \rangle_{\rm ini} = 1.241 \pm 0.017$ and $1.124 \pm 0.009$
for the two reactions. Their difference amounts to $82 \pm 14\%$ of the difference between the fully equilibrated compositions.

\begin{figure} [tbh]
\centerline{\includegraphics[width=8.5cm]{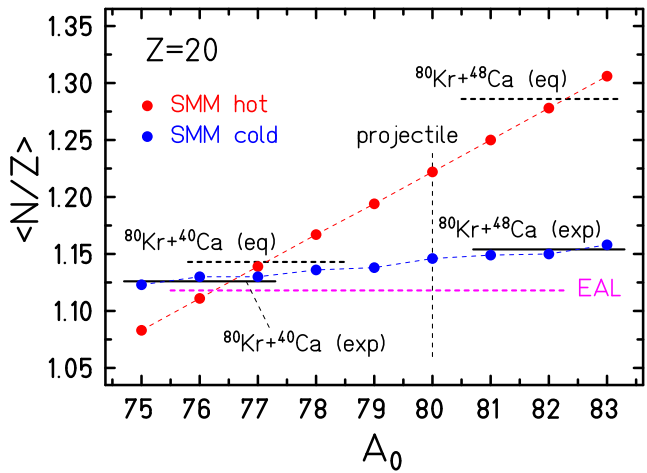}}
%\centerline{\includegraphics[width=8.5cm]{mean_n_over_z_bis.eps}}
\caption{\small{Mean isotopic compositions $\langle N/Z \rangle_{\rm ini}$ of the initial systems (hot, red symbols) and the $\langle N \rangle /Z$
of the final $Z=20$ isotope distributions obtained with the SMM (cold, blue symbols) as a function of the maximum mass $A_0$ of the chosen ensembles.
The calculated results are compared with the experimental mean neutron-over-proton ratios (labeled ``exp", full black lines) for both reactions and 
the EAL prediction (dashed purple line) for $Z=20$ fragments. The compositions corresponding to the isotopically equilibrated projectile and target 
systems (labeled ``eq", dashed black lines) are shown in addition.
}}
\label{fig:nz_mean}
\end{figure}

In addition to the $\chi^2$ analysis employing the experimental errors, the measured and calculated mean mass numbers for each of the six elements were 
used to confirm the obtained results. The example for the same $Z=20$ case and for the two reactions is shown in
Fig.~\ref{fig:nz_mean}. The figure, first of all, documents the strong attenuation of the initial $N/Z$ ratios caused by the breakup and deexcitation of the 
excited sources as modeled with the SMM. Significantly different compositions of the hot sources (red symbols) are required for obtaining the 
small differences of the experimental mean fragment masses. 
The comparison of the experimental $\langle N \rangle/Z$ with the values for cold fragments 
obtained with the SMM (blue symbols), identifies $A_0 = 82$ and $A_0 = 76$ as $A_0$ parameters coming closest to reproducing the  
experimental mean fragment masses for $Z=20$. The corresponding compositions $\langle N \rangle/Z_{\rm ini} = 1.278$ and 1.111 are nearly the same 
as those of the $\chi^2$ analysis (Table~\ref{tab:summary}).
 
 % fig. 11

To find the optimum composition for each element, linear fits to the three calculated mean $\langle N \rangle /Z$ closest to the experimental value
were used and the final result determined with a linear interpolation. The results obtained in this way for the six elements are also shown in 
Table~\ref{tab:summary} (fourth column). The displayed uncertainties are very small, reflecting again the precision of the experimental 
mean neutron numbers. 
For obtaining more realistic uncertainties, the
method used in the $\chi^2$ analysis was adopted. Equal weights were given to the results for each of the elements, 
and the uncertainty of the mean composition determined by dividing the mean squared deviations of the six individual values by $\sqrt{5}$.
The obtained compositions $\langle N/Z \rangle_{\rm ini} = 1.236 \pm 0.016$ and $1.101 \pm 0.022$ for the two reactions are, within their errors, 
compatible with the results obtained with the $\chi^2$ method. The difference amounts to $94 \pm 17\%$ of the difference between the equilibrated systems.
%Combining the results with their errors results in $\Delta \langle N/Z \rangle_{\rm ini} = 0.115 \pm 0.015$,
The weighted average of the results obtained with the two methods corresponds to $87 \pm 11\%$ of the difference between the equilibrated systems.

To arrive at a lower limit for the degree of equilibration following from the present study, a possible effect of the 
symmetry-term coefficient $\gamma$ has to be included. On average, we find that
$\Delta\gamma = +1$~MeV causes $\langle N \rangle$ to decrease by approximately 0.10 and vice versa (cf. Fig.~\ref{fig:sigma}). 
Following Fig.~\ref{fig:nz_mean}, it translates into an uncertainty $\Delta \langle N \rangle/Z \approx 0.005$ 
and $\Delta \langle N/Z \rangle_{\rm ini} \approx 0.031$, equivalent to 22\% of the difference between the equilibrated systems. This  
uncertainty is larger but is only effective if changes of $\gamma$ are different in the two reactions by a certain amount, for example additionally
increased by 1 MeV only in the case of $^{80}$Kr + $^{40}$Ca. A global change of $\gamma$
will only change the vertical position of the line describing the calculated final compositions but not its slope determining the separation of the 
two experimental values on this line. Adopting this possibility as cause for an additional uncertainty leads to a lower bound of 87\% -11\% -22\% = 54\% 
for the degree of equilibration reached in the two reactions.

According to the analysis presented here, rapid $N/Z$ equilibration
during the initial reaction stage seems necessary for reaching the observed final isotopic distributions. 
This conclusion depends crucially on the rate with which the mean $\langle N \rangle /Z$ of fragments
(blue symbols in Fig.~\ref{fig:nz_mean}) increases with $\langle N/Z \rangle_{\rm ini}$, a result of the treatment of secondary decay within the SMM.
We also note that, for both reactions,
the initial compositions are less neutron rich than the equilibrated systems. As we will argue below, it may be caused by the drift of 
neutrons into a neck region of low density which is no longer involved in the disintegration of the projectile residues.

As evident from Fig.~\ref{fig:nz_mean}, the breakup and deexcitation of the hot sources proceeds very differently in the two cases.
There is very little variation of the isotopic composition in the $^{80}$Kr + $^{40}$Ca case. The hot and cold SMM values of $\langle N/Z \rangle$
and the experimental value are all very close to the equilibrium composition and not far from the value of the evaporation attractor line~\cite{charity98}. 
In the case of $^{80}$Kr + $^{48}$Ca, the neutron richness of the excited system disappears to a large extent but, also here, 
the evaporation attractor line is not fully reached. 
The valley of stability attracts the decay chains but they are not long enough to end on the evaporation attractor line. 
Apparently, since the multifragmentation mechanism favors the production of neutron rich 
fragments, the deposited energy is primarily used for the disintegration of the system into smaller fragments instead of neutron evaporation.

\section{Neck Emissions}

In Ref.~\cite{FAZIA21}, the group of coincident light fragments with longitudinal velocities 
smaller than those of the coincident heavy fragments is interpreted as either being emitted
in backward direction from the excited projectile residue or as representing emissions
from the intermediate velocity source, the latter containing products of nucleon-nucleon collisions and from
the decay of the neck forming as the main residues separate~\cite{ditoro06}. In the data analysis, the additional
condition was required that the velocities had to be larger than the center-of-mass velocity.

\vspace{2mm}
\begin{figure} [htb]
\centerline{\includegraphics[width=8.5cm]{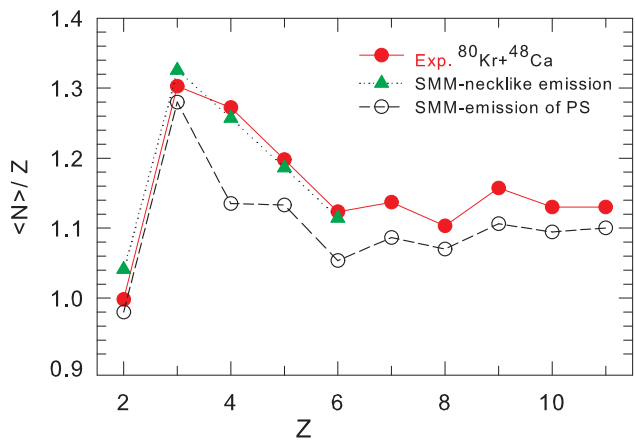}} 
%\centerline{\includegraphics[width=8.5cm]{figures_riza/Figure10_triangle.eps}} 
\caption{\small{Experimental $\langle N\rangle/Z$ values of neck fragments as a function 
of their charge number $Z = 2 - 11$ for the $^{80}$Kr + $^{48}$Ca reaction, in Ref.~\protect\cite{FAZIA21} 
defined as fragments emitted backward with respect to the projectile residue (solid red circles), and 
results of SMM calculations simulating the decay of a neck-type source with $A=19$ and properties given in the text 
for the interval $Z = 2 - 6$ (solid green triangles) in comparison with the results for $Z = 2 - 11$ of the projectile 
source (PS) with $A=81$ and $Z=36$, selected with the requirement of a coincident fragment with $Z \ge 12$ (open circles). 
}}
\label{fig:gonimf}
\end{figure} % fig. 12

\begin{table*}
\caption{\label{tab:calib}
Reconstruction of the composition of the emitting sources obtained with the measured $\langle N \rangle/Z$ ratios of $Z=3,4$ fragments
presented in Figs. 8, 11, and 12(a) of Ref.~\protect\cite{FAZIA21} (third and fifth columns) 
and based on the inclusive $\langle N \rangle/Z$ ratios 
measured for $^{107,124}$Sn fragmentations in experiment S254 (first two lines of Table~\protect\ref{tab:calibz39}). 
For beryllium, the uncorrected $\langle N \rangle/Z$ ratios have been extracted from
the ALADIN data base. The source compositions are obtained by linear interpolations and extrapolations (fourth and sixth columns). Their errors 
represent the effects of the displayed experimental errors reported in Ref.~\protect\cite{FAZIA21}.
}
\begin{ruledtabular}
\begin{tabular}{l c c c c c }
%\hline\hline
& & $Z=3$ (Expt.) & Source & $Z=4$ (Expt.) & Source\\
\hline
group & target & $\langle N \rangle/Z$ & $\langle N/Z \rangle$ & $\langle N \rangle/Z$ & $\langle N/Z \rangle$ \\
\hline
\\
forward & $^{48}$Ca  & $1.255 \pm 0.003$ & $1.29 \pm 0.02$ & $1.180 \pm 0.006$ & $1.38 \pm 0.01$ \\ 
backward & $^{48}$Ca & $1.300 \pm 0.002$ & $1.49 \pm 0.01$ & $1.273 \pm 0.003$ & $1.53 \pm 0.01$ \\ 
forward & $^{40}$Ca  & $1.224 \pm 0.009$ & $1.15 \pm 0.04$ & $1.109 \pm 0.020$ & $1.27 \pm 0.03$ \\ 
backward & $^{40}$Ca & $1.274 \pm 0.016$ & $1.38 \pm 0.07$ & $1.170 \pm 0.032$ & $1.37 \pm 0.05$ \\ 
\\
\end{tabular}
\end{ruledtabular}
\end{table*}             %Table II

The $\langle N \rangle /Z$ ratios for fragments with $Z=2-11$ 
are among the data reported for this group of fragments and, for the $^{80}$Kr + $^{48}$Ca reaction, shown in Fig.~\ref{fig:gonimf}. 
As the authors note, the observed pattern as a function of $Z$ is well
known from other studies and, as an example, the work of Barlini {\it et al.}~\cite{FAZIA13} is cited in~\cite{FAZIA21}. The same characteristic 
odd-even staggering has also been observed in projectile fragmentation at higher energies. The systematics presented in Ref.~\cite{traut07}
for projectiles from nickel to uranium and energies of 600 MeV/nucleon and 1~GeV/nucleon shows that the pattern 
is invariant and that only its location on the $N/Z$ scale depends on the projectile, apparently linearly correlated
with the $N/Z$ ratio of the projectile. In Ref.~\cite{traut07}, a correction for the missing yield of unstable $^8$Be 
was applied to the values for $Z=4$ by including an estimate for it obtained from a smooth interpolation over the 
identified yields of $^{7,9-11}$Be. This correction has a negligible effect for the case of $^{107}$Sn but lowers the 
$\langle N \rangle /Z$ of beryllium for $^{124}$Sn fragmentations from 1.24 to 1.18 which 
makes the systematic odd-even staggering more clearly visible for the neutron rich case. % (cf. Ref.~\cite{ricciardi04} deals with Z>8). 

For the interpretation of the fragmentation reactions at relativistic energies, it may be assumed that the excited sources
inherit the neutron-over-proton ratio of the original projectiles. It is motivated by the reduced nucleon wave lengths, the
dominance of nucleon-nucleon dynamics during the initial reaction stages, and was shown to be successful in applications 
as, e.g., the SMM analysis of the
$^{107,124}$Sn and $^{124}$La fragmentations at 600 MeV/nucleon~\cite{Ogul11}. It implies that the systematics 
obtained for light fragments may be used for calibrating the isotopic compositions of the excited sources. 

In the following, we will assume that the observed correlation between sources and fragments holds for the present 
type of reactions as well.
The statistical decay of excited sources should only depend on their properties and not on how they were formed.
With the measured $\langle N \rangle /Z$ values for products with $3 \le Z \le 9$ from the $^{107}$Sn and $^{124}$Sn 
fragmentations displayed in Fig.~3 of Ref.~\cite{traut07}, a source reconstruction is attempted, not only for
the so-called backward emitted group of light fragments (Fig.~\ref{fig:gonimf}) but also for the 
forward emitted fragments whose $\langle N \rangle /Z$ ratios for $Z=1-4$ are shown in Fig.~8 of~\cite{FAZIA21} and
reproduced in the present Fig.~\ref{fig:14_noverz}. In the latter case, the overlap in atomic number with the $^{107,124}$Sn
fragmentations is restricted to $Z=3$ and 4. The corresponding $\langle N \rangle /Z$ ratios for backward emitted 
fragments from $^{80}$Kr + $^{40}$Ca are obtained by adding the backward-forward differences shown in Fig. 11 of~\cite{FAZIA21}.

The results for the two elements $Z=3$ and 4 are listed in Table~\ref{tab:calib}. They are obtained 
by assuming that the values measured for the $^{107}$Sn and
$^{124}$Sn fragmentations, given in the first two lines of Table~\ref{tab:calibz39}, are representative for source compositions 
$\langle N/Z \rangle =1.16$ and 1.48, respectively, and that a linear interpolation between (or extrapolation from) these values is appropriate. 
The value $\langle N/Z \rangle =1.16$ for the case of $^{107}$Sn refers to the actual
radioactive beams used in the experiment which, besides the desired $^{107}$Sn, contained contributions of neighboring isotopes 
whose intensities had been determined~\cite{Ogul11,lukasik08}. 

As a main conclusion, we observe that the sources responsible for the so-called backward-emitted fragments in both reactions
are significantly more neutron rich than the sources responsible for the forward-emitted fragments. Simply on the basis of the
obtained neutron-over-proton ratios, it can thus be excluded that both groups of fragments have the same origin.
In the case of $^{80}$Kr + $^{48}$Ca, the average composition $\langle N/Z \rangle = 1.51$ of the backward-emitted fragments 
is still significantly 
larger than the value 1.31 expected for a mixed source consisting of equal numbers of nucleons from the projectile
and from the target. The same holds for $^{80}$Kr + $^{40}$Ca. Also here the average composition $\langle N/Z \rangle = 1.37$
largely exceeds the value 1.11 of an equally mixed source. The values
obtained for the forward-emitted particles, on the other hand, are in approximate agreement with the results for the fragments
with $Z=19-24$ shown in Table~\ref{tab:summary}. It permits the conclusion that the latter two groups of 
fragments result from the breakup of the same excited intermediate sources. 

The errors displayed in Table~\ref{tab:calib} are those of the experimental data reported in Ref.~\cite{FAZIA21} and their effect on 
the reconstructed source compositions. These errors are very small. To obtain an estimate of a potential systematic uncertainty 
associated with the method, $\langle N \rangle/Z$ ratios for $Z = 5 - 9$ from the fragmentation of $^{124,136}$Xe projectiles on Pb targets 
are used which had been measured at the GSI Fragment Separator FRS~\cite{henzlova08}. Reconstructing the source
compositions on the basis of the $^{107,124}$Sn results for the same $Z = 5 - 9$ fragments (Table~\ref{tab:calibz39}) leads to values 
consistently higher by $\Delta N/Z = 0.09$ than the $N/Z$ ratios of the $^{124,136}$Xe projectiles, corresponding to almost five additional neutrons. 
The difference $\Delta N/Z = 0.23$ between the results for the two Xe projectiles, however,
is very close to the actual difference $\Delta N/Z = 0.22$ of their $N/Z$ ratios. Relative differences are preserved but
the absolute values, according to this test, are shifted by $\Delta N/Z \approx 0.1$. For the following discussion, the source compositions
listed in the tables will be reduced by $\Delta N/Z = 0.05$ and an error of equal magnitude will be assigned to them. It is equivalent to giving equal weights
to the Sn and Xe fragmentations in their role for the intended reconstruction of source compositions.

\begin{table}
\caption{\label{tab:calibz39}
Reconstruction of the neutron-over-proton ratios 
of the emitting sources (fifth column) obtained with the $\langle N \rangle/Z$ ratios of $Z=3-9$ fragments
presented in Fig. 12 of Ref.~\protect\cite{FAZIA21} (second column) and based on the inclusive $\langle N \rangle/Z$ ratios 
measured for $^{107,124}$Sn fragmentations in experiment S254 as reported in Ref.~\cite{traut07} (third and fourth columns). 
For beryllium, the uncorrected $\langle N \rangle/Z$ ratios have been extracted from
the ALADIN data base. The errors of the sources represent the effects of the displayed experimental errors
reported in Ref.~\protect\cite{FAZIA21}.
}
\begin{ruledtabular}
\begin{tabular}{c c c c c }
%\hline\hline
Z & Expt. & $^{107}$Sn & $^{124}$Sn & Source\\
\hline
\\
3 & $1.300 \pm 0.002$ & 1.227 & 1.297 & $1.494 \pm 0.009$ \\  
4 & $1.273 \pm 0.003$ & 1.038 & 1.243 & $1.527 \pm 0.005$ \\
5 & $1.196 \pm 0.001$ & 1.134 & 1.200 & $1.461 \pm 0.005$ \\
6 & $1.122 \pm 0.001$ & 1.037 & 1.138 & $1.429 \pm 0.003$ \\
7 & $1.137 \pm 0.001$ & 1.103 & 1.153 & $1.378 \pm 0.006$ \\
8 & $1.103 \pm 0.001$ & 1.048 & 1.120 & $1.404 \pm 0.004$ \\
9 & $1.158 \pm 0.002$ & 1.132 & 1.190 & $1.303 \pm 0.011$ \\
\\
\end{tabular}
\end{ruledtabular}
\end{table}              %Table III

This uncertainty does not affect the resulting implications. The so-called backward emitted particles are unlikely to originate from the same
source as the forward emitted particles and light fragments. 
This was also concluded by the FAZIA Collaboration and reported in their paper~\cite{FAZIA21}. 
The neutron richness of the backward sources is enhanced even in comparison to those of the equally mixed 
sources by corrected values $\Delta \langle N/Z \rangle \approx 0.15$ in the case of $^{80}$Kr + $^{48}$Ca and $\approx 0.21$ in the case of
$^{80}$Kr + $^{40}$Ca. If this is caused by neutrons drifting into these sources their densities will have to be low, a property expected
for the neck region located between the separating projectile and target residues~\cite{ditoro06}. Even its mass may be estimated by taking into account that
the compositions of the projectile sources were, on average, lower than those of the equilibrated systems 
by $\Delta \langle N/Z \rangle \approx 0.035$, a change achieved by the drift of two neutrons out of a system with together 56 protons.
The same two neutrons will raise the neutron-over-proton ratio by $\approx 0.18$, the average enhancement observed for the two reactions,
in a smaller system with only about one fifth the number of protons.
The mass of the neck region, according to this estimate, should thus be of the order of $(\langle N/Z \rangle +1) \times 56 \times 0.035/0.18$, i.e.,
about 25 nucleons.

The obtained source properties are very consistent. Going more into detail, one observes that the values for the
sources of $Z=4$ fragments are systematically larger than those for the sources of $Z=3$ fragments. On the other hand,
the calibration for $Z=3$ amplifies small differences and thus may be associated with a larger uncertainty (cf. Table~\ref{tab:calib}).
Elements with odd $Z$ generally respond less strongly to the source composition, according to the data published in Ref.~\cite{traut07} which
extend up to $Z=9$. It is thus possible to extend the calibration to heavier 
fragments of the so-called backward emitted group in $^{80}$Kr + $^{48}$Ca whose composition is presented in the FAZIA
report~\cite{FAZIA21}. 

The results, again obtained with linear interpolations and extrapolations, are given in Table~\ref{tab:calibz39}. 
The obtained compositions are all neutron rich, starting with values $\langle N \rangle/Z \approx 1.5$ for $Z = 3$ and 4 
and slowly decreasing with $Z$ to values 1.4 and 1.3 for the $Z=8$ and 9 fragments, respectively. A dependence on $Z$ of this kind is not unexpected
considering the so-called hierarchy effect according to which the lightest fragments are expected from the regions of lowest density in between
the projectile and target rapidities~\cite{ditoro06,montoya94,colin03}.
The scatter of the individual results may be used as indication of their systematic uncertainty. 

It will finally be shown that SMM calculations of the deexcitation of a neck-type source with properties corresponding
to these results will lead to light fragment yields with $\langle N \rangle /Z$ ratios in good agreement with the FAZIA measurement
for the $^{80}$Kr + $^{48}$Ca reaction~\cite{FAZIA21}.
Following a series of test calculations with varying parameters
for the masses and excitation energies of the source as well as for its $\langle N/Z \rangle$ composition, 
the following scenario was assumed: a fixed mass $A=19$ and atomic number $Z=8$ of the source, corresponding to $N/Z = 1.375$,
an excitation energy of 5 MeV/nucleon, a reduced breakup density $\rho = \rho_0/6$, and a reduced symmetry-term coefficient $\gamma = 14$~MeV
were considered as appropriate for the small masses of the produced fragments.  
Similar parameters were found useful in previous SMM studies of multifragmentation reactions~\cite{SMM,MSU,INDRA}. 
For example, in central $^{197}$Au + $^{197}$Au collisions at the same beam energy of 35 MeV per nucleon,  
a statistical source with the same density and excitation energy per nucleon was extracted~\cite{MSU}.

The obtained $\langle N \rangle /Z$ ratios for neck fragments with $Z \le 6$ are shown in Fig.~\ref{fig:gonimf}, and a very satisfactory agreement 
with the experimental results is observed. For comparison, the results for the projectile source (PS) with $A=81$ and $Z=36$ ($N/Z = 1.25$) is shown as well. 
They are obtained from the same calculation as the results shown in Figs.~\ref{fig:14_48} and \ref{fig:14_noverz} and seen there to agree well
with the experimental results for the forward emitted coincident fragments. The characteristic odd-even staggering is apparent in both calculations as
nuclear structure effects are taken into account in the SMM for the ground-state masses which are restored at the end of the secondary deexcitation.

The significant difference in $\langle N \rangle /Z$ with respect to the experimental values for the class of slower fragments, documented
for $Z=3,4$ in Table~\ref{tab:calib}, is found to extend to larger $Z$. The modest reduction of the difference toward $Z=11$ is not
unexpected. Neck fragments are predominantly small as already mentioned in the discussion related to Table~\ref{tab:calibz39}.
It thus appears that the definition of two groups by the FAZIA Collaboration represented a very 
useful choice, giving evidence for the existence of different sources of coincident light fragments. The sorting according to the sign of the
relative velocity is clearly not sufficient for precisely distinguishing particles and fragments according to their various origins, 
as discussed in detail in the paper~\cite{FAZIA21}.

\section{Discussion} 

The indicated isotopic exchange between projectile and target seems somewhat unexpected on the
basis of the measured isotope distributions exhibiting rather small differences. 
The latter are the consequence of the attraction of the valley of stability which causes the initial neutron richness of the
excited sources in $^{80}$Kr + $^{48}$Ca to quickly disappear (Fig.~\ref{fig:nz_mean}). Large contributions of the target to 
measured isotopic ratios of light fragments have already been observed in the earlier experiment of the FAZIA Collaboration~\cite{FAZIA13}.
At that time, a three-element telescope placed at $\theta_{\rm lab} = 5.4^{\circ}$ was used to measure inclusive fragment production in the
$^{84}$Kr + $^{112,124}$Sn reactions at 35 MeV/nucleon. The observed $\langle N \rangle/Z$ values of light fragments depend significantly on the target.
For lithium and beryllium fragments, values of $\approx 1.3$ are reported with differences of $\approx 0.03$ for the two target cases. 

Applying the same method 
of source reconstruction as for the present reaction leads to rather neutron rich sources with $\langle N/Z \rangle_{\rm ini} = 1.63$ and 1.46 for the reactions with $^{124}$Sn and
$^{112}$Sn targets, respectively. These values are higher than in the present reactions (Tables~\ref{tab:calib} or \ref{tab:calibz39}) as expected
for the more neutron rich projectile. They are also higher than the $\langle N/Z \rangle_{\rm ini} = 1.42$ and 1.28 of the equilibrated $^{84}$Kr + $^{124,112}$Sn
systems and their difference 0.17 exceeds even slightly the difference 0.14 of the latter. 
Following the interpretation offered in the previous section, one would, therefore, conclude that these light 
fragments are primarily emitted from low-density regions of the reaction volume whose neutron content has been enhanced. According to the analysis of
the slightly heavier Xe+Sn reaction by {\L}ukasik {\it et al.}~\cite{lukasik97}, 
the very light fragments have their origin predominantly in the midrapidity regime. 

Isotopic equilibrium is not unusual in the Fermi energy domain.
It was, for example, observed by Johnston {\it et al.} for inclusive reactions of Ar and Ca beams of mass $A=40$ on Fe and Ni targets 
of mass $A=58$ and incident energies 33 MeV/nucleon~\cite{johnston96}. As shown in that work, when the energy is increased to 45 MeV/nucleon
full equilibrium is no longer achieved. Partial equilibrium was observed also in the cross bombardment of $^{112}$Sn and $^{124}$Sn beams and targets at the
slightly higher beam energy of 50 MeV/nucleon by Tsang {\it et al.}~\cite{tsang04}. Partial isospin equilibrium of $\approx 53\%$ was reported by 
Keksis {\it et al.} for $^{40,48}$Ca on $^{112,124}$Sn reactions at the slightly lower energy 32 MeV/nucleon~\cite{keksis10}. 

In the most recent
experiment of the FAZIA Collaboration, isospin transport ratios were determined for cross bombardments of $^{58}$Ni and $^{64}$Ni projectiles and targets at
32 and 52 MeV/nucleon incident energy~\cite{ciampi22}. Increasing isospin equilibration with decreasing impact parameter is reported but full
equilibration is not observed, even for the most central collisions. A very similar conclusion was presented very recently 
for the $^{40,48}$Ca + $^{40,48}$Ca system at 35 MeV/nucleon by Fable {\it et al.} who also applied the isospin transport-ratio technique~\cite{fable23}.
These results emphasize the transitional nature of the Fermi energy domain with the effect that small differences in the reaction systems and the
experimental conditions may manifest themselves in measurable differences of the observed isospin transport~\cite{coupland11}. 

Contact times, if assumed to correspond to the sum of the diameters of the collision partners divided by the velocity of the projectile, are of the
order of 70 fm/$c$ in the present reaction and, depending on the impact parameter and dynamical details, may fall between 50 and 100 fm/$c$. 
This interval represents the timescale on which the observed isospin equilibration apparently occurs. 
It is also consistent with the time characterizing the statistical nucleation process in the neck region which is 
described within the freeze-out volume conception.
Similar times of about 100 fm/$c$ were found by Tsang {\it et al.}~\cite{tsang04} to be needed for the observed isospin diffusion 
in the $^{112,124}$Sn + $^{112,124}$Sn collisions at 50~MeV/nucleon according to Boltzmann-Uehling-Uhlenbeck (BUU) calculations performed for these reactions.
For the similar Xe + Sn system at 32~MeV/nucleon, isospin equilibration times of less than 200~fm/$c$ were deduced from calculations with the stochastic
mean field (SMF) model~\cite{ademard14}.

Equilibration times of very similar magnitude were deduced from an experiment of different type by Jedele {\it et al.}~\cite{jedele17}. There, the authors 
followed the isospin equilibration
in a rotating dinuclear system formed by the decay of deformed projectile fragments produced in $^{70}$Zn + $^{70}$Zn collisions at 
35 MeV/nucleon. 
For selected systems, the obtained equilibration times %consisting of nuclei with atomic numbers $Z=12-14$ and $Z=5-7$ was found to be 
were of the order of 0.3 zs, corresponding to $\approx 100$ fm/$c$, whereas the process up to full isospin equilibration may be followed
for somewhat longer times~\cite{brown13}.
It thus seems that, 
according to the presented examples, $N/Z$ equilibration proceeds fast enough to be observed in the $^{80}$Kr + $^{48,40}$Ca experiments
at 35 MeV/nucleon. It should be interesting to study the full charge range of projectile fragments and connect it with the evolution 
of the isospin transport. The measurements of correlated fragments, 
both in the neck and projectile regions, are important in this case.

The neutron richness of neck emissions has been reported by many authors~\cite{lukasik97,fable23,theriault06,kohley11,Filip12,bougault18} and deduced from
theoretical work~\cite{baran04}. Th\'{e}riault {\it et al.} have studied the symmetric $^{64}$Zn + $^{64}$Zn reaction at 45 MeV/nucleon~\cite{theriault06}.
They found that the completely reconstructed midrapidity material is more neutron rich than the overall reaction system whereas 
the reconstructed $N/Z$ ratio for the projectile system remains below that value. This result was presented by the authors as the most complete evidence, 
at that time, of neutron enrichment of midrapidity nuclear matter at the expense of the remaining reaction system, 
a phenomenon related to the density dependence of the nuclear symmetry energy. 
The reconstruction of the compositions of the projectile and neck source presented here for the $^{80}$Kr + $^{48,40}$Ca reactions 
is in full agreement with these earlier findings.

\section{Summary}

The isospin transport in the $^{80}$Kr + $^{48}$Ca and $^{80}$Kr + $^{40}$Ca reactions at 35 MeV/nucleon was studied with calculations using the
statistical multifragmentation model (SMM, Ref.~\cite{SMM}) in comparison with the data reported by the FAZIA Collaboration~\cite{FAZIA21}.
Regarding the experimental data, it was assumed that the events recorded with four FAZIA blocks are representative for the mid-peripheral collisions
of the studied reactions. In the analysis, the ensembles of excited sources formulated and verified in previous studies of ALADIN and other experimental data
were considered as applicable for the present reactions, an
assumption supported by the obtained very satisfactory representation of the measured isotopic distributions. The odd-even staggered $\langle N\rangle /Z$
pattern of light fragments and its relation to the composition of the source was assumed to be universally valid.

With these assumptions, the intermediate reaction system at the time of the transition from its dynamical creation to the assumed statistical decay
was reconstructed by identifying it with the ensemble of sources whose decay is treated with the SMM. It was found that the isospin exchange
leads the intermediate system to cover at least half of the way to equilibrium. % with full equilibrium not being excluded.
The most probable value of $87 \pm 11\%$, in fact, does not exclude full equilibrium. 
The reconstructed compositions were found to be slightly reduced in neutrons with respect to the equilibrium compositions
by amounts corresponding to a loss of about two neutrons, similar for both reactions. 

The compositions of the sources responsible for the production
of the two groups of light fragments with $Z \le 4$ were found to be significantly different. 
For the group with longitudinal velocities larger than those of the coincident heavier fragments, 
the source composition is similar to that of the heavier fragments, indicating that both, 
the heavier and the lighter fragments, originate from the breakup of the same intermediate system. The sources of light fragments with longitudinal velocities
smaller than those of the coincident heavier fragments were found to be enriched in neutrons beyond the values
expected for mixed systems consisting of equal numbers of nucleons from the projectile and from the target. The assumption of a low-density neck region,
enriched in neutrons and with a mass of the order of 25 nucleons, provided a consistent explanation. 

These conclusions are based on SMM calculations using ensembles of excited sources as initial configurations whose decay is assumed to be statistical
and followed according to the model. The simultaneous consideration of two correlated statistical sources, 
one representing the projectile residues and one describing the excited low-density matter in the neck region, represents a novel
theoretical development. 
The applicability of the method was demonstrated by satisfactorily reproducing the reported isotope distributions with reasonable assumptions 
regarding the effect of the target composition in forming the excited projectile residues. 

In the second part of the analysis, the isotopic composition of the
ensemble was systematically varied and the agreement of the calculated and experimental results quantitatively evaluated. Two methods were used,
one of comparing the experimental and theoretical mean $\langle N \rangle/Z$ values, and the other of minimizing the $\chi^2$ of the reproduction
of the measured isotope distributions by using the experimental errors. 
Both methods were found to yield consistent results. 
The $\langle N \rangle/Z$ pattern as a function of $Z$ of fragments from the neck-type source was successfully described
by assuming a lower density and higher excitation energy. 
These parameters are similar to those found in the statistical analyses of central 
nucleus-nucleus collisions at similar energies. Altogether, the chosen statistical approach using two correlated sources,
was found applicable for studying isospin transport phenomena in mid-peripheral collisions in the Fermi energy domain.

\begin{acknowledgments}
The authors are grateful to S. Piantelli and the FAZIA Collaboration for providing 
published data in numerical form
and to the ALADIN Collaboration for giving access to the S254 experimental data.
R.O. gratefully acknowledges the support by TUBITAK (Turkish Scientific 
and Technological Research Council) under Project No. 
BIDEB-1059B192001096 and the warm hospitality of GSI. 
N.B., A.S.B., and M.B. acknowledge support from 
the DAAD/TUBITAK exchange program.
\end{acknowledgments}

\end{document}